\documentclass[aps,prd,groupedaddress,showpacs,notitlepage]{revtex4-2}
\usepackage{amsmath,amstext,amsbsy,amssymb}
\usepackage{bm}
\usepackage{color}

\usepackage[linktocpage]{hyperref}

\newcommand{\diag}{{\rm diag}}

	\newcommand{\ssR}{{\scriptscriptstyle{ R}}}

	\newcommand{\ssL}{{\scriptscriptstyle{ L}}}

	\long\def\symbolfootnote[#1]#2{\begingroup%
		\def\thefootnote{\fnsymbol{footnote}}\footnote[#1]{#2}\endgroup}

\begin{document}
		
		\title{
			Lorentz-violating extension of    Wigner function formalism and chiral kinetic theory
		%	\\   Lorentz-violating extension of  the quantum kinetic equation and chiral transport  
		}
		
		\author{\"{O}mer F. Dayi}
	\email{dayi@itu.edu.tr } 
\affiliation{%
	Physics Engineering Department, Faculty of Science and
	Letters, Istanbul Technical University,
	TR-34469, Maslak--Istanbul, Turkey}

\begin{abstract}
The quantum kinetic equation for the gauge-invariant Wigner function, constructed from spinor fields that obey the Dirac equation modified by $CPT$ and Lorentz symmetry-violating terms, is presented. The equations for the components of the Wigner function in the Clifford algebra basis  are accomplished.
Focusing on the massless case, an extended semiclassical chiral kinetic theory in the presence of external electromagnetic fields is developed. We calculate the chiral currents and establish the anomalous magnetic and separation effects in a Lorentz-violating background. The chiral anomaly within the context of extended quantum electrodynamics is elucidated. Finally, we derive the semiclassical Lorentz-violating extended chiral transport equation.
\end{abstract}

\maketitle
	
\section{Introduction}

The fermionic constituents of  the Standard Model of particle physics are mainly the  charged Dirac and Weyl particles which also appear   in condensed matter as quasiparticles.  Various methods have been employed to study the dynamical features of spin-1/2 fermions. 
Nevertheless, an   intuitive understanding of  physical phenomena involving them is offered by kinetic theory.
A systematic  construction of  kinetic theory begins with the quantum kinetic equation  satisfied by the relativistic Wigner function  \cite{egv,vge}. It is the relativistic quantum field theory  counterpart of the classical distribution function. In high-energy physics, quantum kinetic theory has predominately been investigated  to elucidate phenomena arising in the  heavy-ion collisions due to the chiral nature of quarks in a novel phase of quark-gluon plasma.   A  comprehensive review of quantum transport theory derived by means of the Wigner function method and its applications is provided  in  \cite{HidakaRev}.

One of the main properties of quantum field theory models is the invariance under  Lorentz transformation. However,
there are debates about the violation of Lorentz symmetry under certain conditions, such as very high energies.  For instance, it may
arise from the string theory as  outlined in \cite{PhysRevD.39.683, KPString}.   On the other hand, Lorentz symmetry violation  may stem from  anisotropy in spacetime.
A brief historical overview and current status of ideas regarding potential sources of Lorentz violation in particle physics can be found in \cite{Jose}.
Our discussion is based on the Standard Model extension (SME), an effective field theory that incorporates Lorentz- and $CPT$-violating terms  into the Standard Model of particle physics \cite{CollKost,PhysRevD.58.116002}. The SME assumes that a fundamental theory, which may be string theory, undergoes spontaneous Lorentz symmetry breaking. This results in a low-energy effective action with explicit Lorentz symmetry-violating (LSV) terms while maintaining microcausality, positivity of energy, and energy-momentum conservation.
Notably, in SME  particle and observer Lorentz transformations are not treated  on equal footing.
Observer frames differ in orientation and velocity, hence they are related to coordinate changes.
The fundamental theory's Lorentz symmetry is  spontaneously broken, so that  observer Lorentz symmetry  is conserved. However, particle Lorentz symmetry is broken because it is defined by rotations and boosts of the localized field while keeping the expectation values of tensor fields unchanged.
Restricting the minimal SME to Abelian gauge theory one obtains  the extended quantum electrodynamics (QED) whose fermionic sector is given by the Dirac spinors coupled to electromagnetic gauge fields in the presence of Lorentz- and $CPT$-violating terms. This framework is also utilized to discuss violation of emergent Lorentz symmetry in Dirac and Weyl semimetals \cite{DWSM}.  

We  deal with the modified  Dirac equation with a certain set of LSV terms: $a_\mu$ and $c_{\mu\nu}.$ The Boltzmann equation for this choice of LSV coefficients has been examined in \cite{Potting}. The spinor fields obeying the modified Dirac equation can be quantized using the customary methods of field theory, and 
the  extended QED is invariant under the gauge transformations.  Thus,  one can construct  the extended relativistic gauge-invariant  Wigner function and derive
the quantum kinetic equation  as in the ordinary relativistic kinetic theory.  
We expand  the Wigner function in the Clifford algebra basis where the coefficients  are scalar, pseudoscalar, vector, axial-vector,  and tensor field variables.  The quantum kinetic equation yields the coupled equations of  these fields, which simplify when  the mass  is set to zero.  
We then consider the massless particles and  explore the chiral kinetic theory which is manifestly observer Lorentz  invariant. One of the peculiar aspects of chiral fermions is the emergence of   anomalous effects like  chiral magnetic \cite{kmw,fkw}  and separation \cite{kz, mz, jkr} effects.   By choosing the distribution function appropriately we will be able to calculate vector and axial-vector particle currents and show that these anomalous effects get contribution from the LSV terms. Moreover, by studying their conservation we will demonstrate that the chiral anomaly is also  altered.

In the next section we discuss the relativistic gauge-invariant  Wigner function and the quantum kinetic equation for Dirac particles within extended QED. In Sec. \ref{cvf} the quantum kinetic equations of the chiral vector fields  will be presented. Two of the three coupled equations are solved  in a semiclassical approach.  
In Sec. \ref{FDD} we employ the modified Fermi-Dirac distribution function   \cite{Potting}  and calculate the chiral vector fields, which are then used to establish four-currents generating anomalous magnetic and separation effects. The vector current is shown to be conserved, while the axial-vector four-current leads to the chiral anomaly. In Sec. \ref{kinEq}, an observer Lorentz-invariant semiclassical chiral kinetic equation is derived. It is followed by a discussion of results and potential future studies in the final section.

%\begin{equation}{\cal L}_{LSV}=\bar{\psi}(x)	\left[ \Gamma_{L}^\mu  (i\hbar \partial_\mu - q A_\mu)- M_L \right] \psi (x) ,\label{GdirL}\end{equation} where \begin{eqnarray}\Gamma_{L}^\nu &= &\gamma^\nu +c^{\nu\mu}\gamma_\mu +d^{\mu \nu}\gamma_5\gamma_\mu +e^\nu +if^\nu \gamma_5 +\frac{1}{2}g^{\lambda \mu \nu} \sigma_{\lambda \mu}  ,\nonumber \\M_L & =& m +a_\mu \gamma^\mu +b_\mu \gamma_5 \gamma^\mu +\frac{1}{2} H^{\mu\nu} \sigma_{\mu\nu}.\end{eqnarray}Here  we deal with the fermion sector of the extended QED by choosing $d=e=f=g=b=H=0.$ 

\section{Quantum kinetic equation}
 \label{QKE-Sec}

We deal with  the Dirac fermion of mass $m$ and charge  $q$ in the presence of  external electromagnetic fields. There may be several Lorentz symmetry-violating terms that are coordinate reparametrization  and gauge invariant \cite{KostLehn}. However, we consider a restricted set of Lorentz symmetry-violating coefficients 
 described by
the Lagrangian density of the Dirac spinors $\psi$ coupled to  the electromagnetic gauge field $A_\mu,$  given as $(c=1)$
\begin{equation}
{\cal L}=\bar{\psi}(x)	\left[ \Gamma^\mu  (i\hbar \partial_\mu - q A_\mu)- M \right] \psi (x) ,
	\label{dirL}
\end{equation}
where $\partial_{\mu} \equiv \partial / \partial x^\mu$   and 
\begin{equation}
	\Gamma_\mu = \gamma_\mu +c_{\nu \mu } \gamma^\nu,\ \ \ M=m +a_\mu \gamma^\mu . \nonumber
\end{equation}
The Minkowski metric is $g_{\mu \nu}=\diag (1,-1,-1,-1),$ and $\gamma_\mu$ are the ordinary $\gamma$  matrices satisfying $\{\gamma_\mu ,\gamma_\nu\}=2g_{\mu \nu}.$ The LSV coefficients
 $a_\mu$ and $c_{\mu\nu}$ are  real and  constant. The former violates also the $CPT$ invariance.   
 By varying (\ref{dirL})  with respect to $\bar{\psi}$ and $\psi ,$ one derives 
the equations of motion as
\begin{eqnarray}
\left[\Gamma^\mu  (i\hbar \partial_\mu - q A_\mu)-M \right] \psi (x) &=& 0,
\label{dirP} \\
 \bar{\psi} (x)\left[\Gamma^\mu  (i\hbar \partial^\dagger_\mu  +q A_\mu)+M \right] &=& 0. \label{dirPB} 
\end{eqnarray}

In \cite{CollKost, KostLehn} it has been demonstrated  that  the spinor operators can be defined as in the ordinary case:  In the ``concordant frame'' where the LSV coefficients are small, there exists a transformation of the spinors $\psi (x)$ 
that leads to a free Hamiltonian possessing two positive and two negative eigenvalues.
 Hence, one introduces plane wave solutions and define wave packets. Then, one proceeds as in the ordinary quantum field theory  and introduces the spinor operators $\hat{\psi}(x)$ and  $\hat{\bar{\psi}}(x).$ Therefore, by means of the four-momentum $p^\mu$  one can define the Wigner operator as
$$
\hat{W}_{\alpha \beta} (x,p)=\int \frac{d^4 y}{(2\pi \hbar)^4}e^{-ip\cdot y/\hbar}  \bar{\hat{\psi}}_\beta  (x+\frac{1}{2}y )  \hat{\psi}_\alpha (x-\frac{1}{2}y ).
$$ 
By normal ordering $(:\, :)$ and  ensemble averaging $(\langle \cdots \rangle)$  the Wigner operator, one acquires  the  Wigner function  \mbox{$W (x,p)=\langle : \hat{W} (x,p) : \rangle .$} 
In the presence of electromagnetic interactions one can introduce the gauge-invariant Wigner function 
by means of the gauge link as in the ordinary formalism \cite{egv,vge}. The derivation of the quantum kinetic equation satisfied by the Wigner function relies only on the Dirac equations. Therefore,  employing  the modified Dirac equations (\ref{dirP}) and (\ref{dirPB}), it is accomplished as follows:
\begin{equation}
	\left[\Gamma_\mu  K^\mu-M \right] W(x,p) = 0,
	\label{qke}
\end{equation}
 where
\begin{equation}
\label{kmu}
	K^\mu = \left(\pi^\mu + \frac{i\hbar}{2} 	{\mathcal D}^\mu \right).
\end{equation} 
We defined
\begin{eqnarray}
	{\mathcal D}^{\mu} &\equiv & \partial^{\mu}-qj_{0}(\Delta) F^{\mu\nu}   \partial_{p \nu} , \nonumber\\
	\pi^{\mu} &\equiv &p^{\mu}-\frac{\hbar q}{2} j_{1}(\Delta)F^{\mu\nu} \partial_{p \nu} , \nonumber
\end{eqnarray}
where  $ \partial_{p}^{\mu} \equiv \partial / \partial p_\mu ,$ and $j_{0}(\Delta),$  $j_{1}(\Delta)$ are the spherical Bessel functions in $\Delta \equiv \partial_p^\mu  \partial_\mu .$
Written in  the Clifford algebra basis the Wigner function becomes
\begin{equation}
	W=\frac{1}{4}\left(\mathcal{F}+i \gamma^{5} \mathcal{P}+\gamma^{\mu} \mathcal{V}_{\mu}+\gamma^{5} \gamma^{\mu} \mathcal{A}_{\mu}+\frac{1}{2} \sigma^{\mu \nu} \mathcal{S}_{\mu \nu}\right).
	\label{wigner}
\end{equation}
Let us choose $
	c_{\mu \nu} =c_{\nu\mu}
$
without loss of generality and  introduce 
\begin{equation}
	G_{\mu\nu}= g_{\mu\nu}+c_{\mu\nu} =G_{\nu\mu},
\end{equation}
which is a  nondiagonal metric \cite{colMc}.

One can show that 
(\ref{qke}) leads to the following set of coupled equations
\begin{eqnarray}
	G^{\lambda \mu}K_\lambda \mathcal{V}_\mu-m \mathcal{F} -a^\mu \mathcal{V}_\mu &=&0, \label{eq1} \\
	i		G^{\lambda \mu}K_\lambda \mathcal{A}_\mu+m \mathcal{P} +a^\mu \mathcal{A}_\mu &= &0, \label{eq11} \\
G_{\lambda \mu} K^\lambda \mathcal{F}-i
G^{\nu \lambda } K_\lambda
\mathcal{S}_{\nu \mu}-m \mathcal{V}_{\mu}-a_\mu \mathcal{F}&=&0,  \label{eq2} \\
i	G_{\lambda \mu}K^\lambda \mathcal{P}+\frac{1}{2} \epsilon_{\mu \nu \alpha \beta} 	G^{\nu \lambda }K_\lambda 
S^{\alpha \beta}-m \mathcal{A}_{\mu}-a_\mu \mathcal{P} &=&0, \label{eq3} \\
i	G_{\lambda \mu}K^\lambda  \mathcal{V}_{\nu} -i	G_{ \lambda \nu }K^\lambda  \mathcal{V}_{\mu} - \epsilon_{\mu \nu \rho \sigma}  	G^{\rho\lambda}
K_\lambda \mathcal{A}^{\sigma}
& &\nonumber \\
 -m \mathcal{S}_{\mu \nu}
-i\left(a_\mu  \mathcal{V}_{\nu} - a_\nu  \mathcal{V}_{\mu} \right) -  \epsilon_{\mu \nu \rho \sigma}   a^\sigma \mathcal{A}^\sigma & =&0. \label{eq4}
\end{eqnarray}
In the subsequent sections we consider the chiral fermions in  the semiclassical  limit.

\section{Chiral vector fields}
\label{cvf}

Observe  that for $m=0$ the equations of the fields $\mathcal{V}_\mu ,\mathcal{A}_\mu ,$ (\ref{eq1}),(\ref{eq11}),(\ref{eq4}), decouple from the rest:
\begin{eqnarray}
	G^{\lambda \mu}K_\lambda \mathcal{V}_\mu-a^\mu \mathcal{V}_\mu &= &0, \label{eqm1} \\
	i	G^{\lambda \mu}K_\lambda \mathcal{A}_\mu+a^\mu \mathcal{A}_\mu  &=&0, \label{eqm11} \\
	i	G_{\lambda \mu}K^\lambda  \mathcal{V}_{\nu} -i	G_{\nu \lambda }K^\lambda  \mathcal{V}_{\mu} - \epsilon_{\mu \nu \rho \sigma}  	G^{\rho\lambda} K_\lambda \mathcal{A}^{\sigma}
	& &\nonumber \\
	-i\left(a_\mu  \mathcal{V}_{\nu} - a_\nu  \mathcal{V}_{\mu} \right) -  \epsilon_{\mu \nu \rho \sigma}   a^\sigma \mathcal{A}^\sigma & =&0.  \label{eqm111}
\end{eqnarray}
In fact, we will deal with the vector field $\mathcal{V}_\mu $ and the axial-vector field $\mathcal{A}_\mu $ for  $m=0$ in the semiclassical limit,  keeping
 at most the $\hbar$-dependent terms. Thus, in terms of 
\begin{eqnarray}
	D^{\mu} &\equiv & \partial_{x}^{\mu}-qF^{\mu\nu}  \partial_{p \nu}  , \label{dmu}
\end{eqnarray}
(\ref{kmu}) becomes
\begin{eqnarray}
	K_\mu &=&p_\mu -\frac{i\hbar}{2} D_\mu \nonumber .
\end{eqnarray}
Now, by introducing
\begin{eqnarray}
\bar{p}^\mu &= & G^{\lambda \mu}p_\lambda -a^\mu  , \label{pmu}\\
	\bar{D}^\mu &=& G^{\lambda \mu}D_\lambda ,
\end{eqnarray}
the real parts of (\ref{eqm1})-(\ref{eqm111}) can be expressed as
\begin{eqnarray}
	 	\bar{p}\cdot \mathcal{V}=0,  \label{real1} \\
	 	\frac{\hbar}{2} \bar{D} \cdot \mathcal{A} =0, \\
\frac{\hbar}{2} \left(	\bar{D}_{\mu} \mathcal{V}_{\nu}-  \bar{D}_{\nu} \mathcal{V}_{\mu} \right)-\epsilon_{\mu \nu \alpha \beta} \bar{p}^{\alpha} \mathcal{A}^{\beta}=0. \label{real4} 
\end{eqnarray}
Moreover, the imaginary parts of (\ref{eqm1})-(\ref{eqm111})  are written as
\begin{eqnarray}
	{\hbar \bar{D} \cdot \mathcal{V}=0}, \label{imag1} \\ 
	{\bar{p} \cdot \mathcal{A}=0}, \label{imag2} \\ 
	\bar{p}_{\mu} \mathcal{V}_{\nu}-\bar{p}_{\nu} \mathcal{V}_{\mu}+\frac{\hbar}{2} \epsilon_{\mu \nu \alpha \beta} \bar{D}^{\alpha} \mathcal{A}^{\beta}=0.
	\label{imag5}
\end{eqnarray}
These two sets of coupled equations can be unified by launching  the chiral vector fields 
\begin{equation}
	\mathcal{J}_\chi=\frac{1}{2} \left(\mathcal{V} +\chi \mathcal{A}\right),
\end{equation}
where $\chi =\pm$ labels the right- and left-handed vector fields.
Then (\ref{real1})-(\ref{imag5}) can be combined into
\begin{eqnarray}
\bar{p} \cdot \mathcal{J}_\chi &=& 0, \label{qke1} \\ 
 \bar{D} \cdot \mathcal{J}_\chi &=& 0, \label{qke2} \\ 
\frac{\hbar}{2} \epsilon_{\mu \nu \alpha \beta} \bar{D}^{\alpha} \mathcal{J}_\chi^{\beta}  &=& -\chi \left( \bar{p}_{\mu} \mathcal{J}_{\chi \nu} -\bar{p}_{\nu} \mathcal{J}_{\chi \mu} \right).	\label{qke3}
\end{eqnarray}
We would like to solve these  equations  by expanding the chiral vector fields in $\hbar$ as $\mathcal{J}_\chi=\mathcal{J}^{(0)}_\chi +\hbar \mathcal{J}^{(1)}_\chi.$ Although in the ordinary case these solutions have been discussed in  several papers like  \cite{glpww, hpy95, hpy97},  a  comprehensive presentation can be found in  \cite{hsjlz}. Therefore,  we mainly follow the formulation of \cite{hsjlz}.  The zeroth order solution of  (\ref{qke1}) and (\ref{qke3})   can easily be identified as 
\begin{equation}
\mathcal{J}^{(0)}_{\chi\mu}= \bar{p}_{\mu} f_\chi^{(0)} \delta (\bar{p}^2) ,
\label{jj0}
\end{equation}
where
$ f_\chi^{(0)} $ is a general distribution function. Equation
(\ref{jj0})  should also satisfy  (\ref{qke2}),
\begin{eqnarray}
\bar{D}^{\mu}\left[\bar{p}_{\mu} f_\chi^{(0)} \delta (\bar{p}^2)\right] = G^{\mu\lambda } \left(\partial_{x\lambda}-qF_{\lambda\kappa}  \partial_{p}^\kappa  \right) \left[\left(G_{\mu\sigma} p^\sigma -a_\mu\right) f_\chi^{(0)} \delta (\bar{p}^2)\right] \nonumber \\
=qG^{\mu\lambda } F_{\lambda\kappa} G^{\kappa}_\mu  f_\chi^{(0)} \delta (\bar{p}^2) 
 +2qG^{\mu\lambda }\bar{p}_{\mu} \bar{p}_{\alpha} F_{\lambda\kappa} G^{\alpha \kappa}  f_\chi^{(0)} \delta^\prime (\bar{p}^2) 
+\bar{p}_{\mu}	\delta (\bar{p}^2) \bar{D}^{\mu} f_\chi^{(0)}  =0. \label{zoq}
\end{eqnarray}
It is convenient to define 
\begin{equation}
	\bar{F}^{\mu\nu} =G^{\mu\alpha}F_{\alpha\beta}G^{\beta\nu},  
\end{equation}
which is antisymmetric
\begin{equation}
 \bar{F}^{\nu\mu} =G^{\nu\alpha}F_{\alpha\beta}G^{\beta\mu}= G^{\alpha\nu}F_{\alpha\beta}G^{\mu\beta} =-G^{\mu\beta}F_{\beta \alpha}G^{\alpha\nu} = -	\bar{F}^{\mu\nu} .
\end{equation}
The first and the second terms in (\ref{zoq}) vanish due to
$
\bar{F}^{\mu}_\mu =0,$ and $\bar{p}_{\mu} \bar{p}_{\nu} \bar{F}^{\mu\nu}=0.$
Hence it yields the modified Vlasov equation
\begin{equation}
\bar{p}_{\mu}	\delta (\bar{p}^2) \bar{D}^{\mu} f_\chi^{(0)}  = \bar{p}_{\mu}	\delta (\bar{p}^2) 	G^{\mu\lambda} \left(\partial_{x\lambda}-qF_{\lambda\kappa}  \partial_{p}^\kappa  \right)  f_\chi^{(0)}=0 \label{Vlasov} .
\end{equation}
At the $\hbar$ order  (\ref{qke1})-(\ref{qke3}) yield
\begin{eqnarray}
	\bar{p} \cdot \mathcal{J}^{(1)}_\chi &=& 0, \label{qke12} \\ 
	\bar{D} \cdot \mathcal{J}^{(1)}_\chi &=& 0, \label{qke22} \\ 
	 \epsilon_{\mu \nu \alpha \beta} \bar{D}^{\alpha} \mathcal{J}^{(0)\beta}_\chi  &=& -2\chi \left( \bar{p}_{\mu} \mathcal{J}^{(1)}_{\chi \nu} -\bar{p}_{\nu} \mathcal{J}^{(1)}_{\chi\mu} \right). \label{f3}
\end{eqnarray}
To solve (\ref{f3}),  note that
\begin{eqnarray}
	\bar{D}^{\alpha}\left[\bar{p}^{\beta} f_\chi^{(0)} \delta (\bar{p}^2)\right] = \delta (\bar{p}^2) \bar{p}^\beta \bar{D}^\alpha  f_\chi^{(0)} -q\bar{F}^{\alpha\beta}      f_\chi^{(0)} \delta (\bar{p}^2)  -2 q\bar{F}^{\alpha\rho}  {\bar{p}}_\rho  \bar{p}^\beta  f_\chi^{(0)} \delta^\prime (\bar{p}^2) . \nonumber 
\end{eqnarray}
By making use of the Schouten identity
\begin{equation}
	k_\mu \epsilon_{\nu\rho\sigma\lambda}+ k_\nu \epsilon_{\rho\sigma\lambda \mu}+k_\rho \epsilon_{\sigma\lambda\mu\nu}+k_\sigma \epsilon_{\lambda \mu \nu\rho}+k_\lambda  \epsilon_{\mu\nu\rho\sigma}=0, \label{sid}
\end{equation}
and defining the dual of $\bar{F}$ as
\begin{equation}
	\tilde{\bar{F}}_{\mu\nu} =\frac{1}{2}\epsilon_{\mu\nu\rho\sigma} \bar{F}^{\rho\sigma},
\end{equation}
we get
\begin{eqnarray}
-	 \epsilon_{\mu \nu \alpha \beta}\bar{F}^{\alpha\rho}  {\bar{p}}_\rho  \bar{p}^\beta =- 2\tilde{\bar{F}}_{\nu\beta} \bar{p}_\mu\bar{p}^\beta -2 \tilde{\bar{F}}_{\beta\mu} \bar{p}_\nu\bar{p}^\beta 
+ \bar{p}_\alpha \epsilon_{\beta\rho\mu\nu} \bar{F}^{\alpha\rho}  \bar{p}^\beta -2 \tilde{\bar{F}}_{\mu\nu} \bar{p}^2 . \label{39}
\end{eqnarray}
The third term on the right-hand side is equal to the term on the left-hand side  up to a minus sign, thus (\ref{39}) yields 
\begin{eqnarray}
\epsilon_{\mu \nu \alpha \beta}\bar{F}^{\alpha\rho}  {\bar{p}}_\rho  \bar{p}^\beta  f_\chi^{(0)} \delta^\prime (\bar{p}^2)
	=(\tilde{\bar{F}}_{\nu\beta} \bar{p}_\mu + \tilde{\bar{F}}_{\beta\mu} \bar{p}_\nu ) \bar{p}^\beta f_\chi^{(0)} \delta^\prime (\bar{p}^2)
	- \tilde{\bar{F}}_{\mu\nu}   f_\chi^{(0)} \delta(\bar{p}^2),
\end{eqnarray}
where we employed the identity $\delta^\prime (\bar{p}^2) =-\delta(\bar{p}^2)/\bar{p}^2.$ Therefore, the left-hand side  of  (\ref{f3}) is expressed as
\begin{eqnarray}
 \epsilon_{\mu \nu \alpha \beta} \bar{D}^{\alpha} \mathcal{J}^{(0)\beta}_\chi = \epsilon_{\mu \nu \alpha \beta}
\delta (\bar{p}^2) \bar{p}^\beta \bar{D}^\alpha  f_\chi^{(0)} 	-2q(\tilde{\bar{F}}_{\nu\beta} \bar{p}_\mu + \tilde{\bar{F}}_{\beta\mu} \bar{p}_\nu ) \bar{p}^\beta f_\chi^{(0)} \delta^\prime (\bar{p}^2). \label{lhs3rd}
\end{eqnarray}
Now, we plug (\ref{lhs3rd}) into (\ref{f3}) and multiply it with $\bar{p}^\nu ,$
\begin{eqnarray}
		-2\bar{p}^2( \tilde{q\bar{F}}_{\beta\mu} \bar{p}^\beta f_\chi^{(0)} \delta^\prime (\bar{p}^2)
-2\chi  \mathcal{J}^{(1)}_{\mu})=0.
\end{eqnarray}
Its general solution can be written as
\begin{eqnarray}
 \mathcal{J}^{(1)}_{\mu}= \chi q \tilde{\bar{F}}_{\beta\mu} \bar{p}^\beta f_\chi^{(0)} \delta^\prime (\bar{p}^2) + \bar{p}_\mu f_\chi^{(1)}  \delta (\bar{p}^2)+H_\mu  \delta (\bar{p}^2) ,
\end{eqnarray}
where  $f_\chi^{(1)} $ is a general distribution function and $H_\mu$ is a vector field satisfying
\begin{eqnarray}
\bar{p}^\mu H_\mu  \delta (\bar{p}^2) =0 \label{pK0},
\end{eqnarray}
and
\begin{eqnarray}
	\epsilon_{\mu \nu \alpha \beta}
\delta (\bar{p}^2) \bar{p}^\beta \bar{D}^\alpha  f_\chi^{(0)} 	 \delta (\bar{p}^2)
	&=& -2\chi \left( \bar{p}_{\mu} H_{\nu} -\bar{p}_{\nu} H_{\mu} \right)  \delta (\bar{p}^2).  \label{eqK}
\end{eqnarray}
By introducing the four-vector $n_\mu ,$ which is defined to satisfy $n^2=1,$ one can solve (\ref{pK0}) and (\ref{eqK}) as
\begin{eqnarray}
H_{\mu}  =\frac{\chi}{2\bar{p}\cdot n}\epsilon_{\mu \nu \alpha \beta}
\bar{p}^\nu n^\alpha \bar{D}^\beta  f_\chi^{(0)} 	.
\end{eqnarray}
By means of the Schouten identity (\ref{sid}) and the modified  Vlasov equation  (\ref{Vlasov}), we can show that  it indeed  satisfies (\ref{eqK}),
\begin{eqnarray}
  \frac{1}{\bar{p}\cdot n} \left( \bar{p}_\nu \epsilon_{\mu \rho \alpha \beta} 
	-\bar{p}_\mu\epsilon_{\nu \rho \alpha \beta} \right) 
\bar{p}^\rho n^\alpha  \delta (\bar{p}^2) \bar{D}^\beta  f_\chi^{(0)} 	& = &
\frac{-1}{\bar{p}\cdot n} \big(
\bar{p}_\mu \epsilon_{ \rho \alpha \beta \nu} + \bar{p}_\rho \epsilon_{ \alpha \beta \nu \mu}  + \bar{p}_\alpha \epsilon_{ \beta \nu \mu \rho } \nonumber \\
&&+ \bar{p}_\beta \epsilon_{  \nu \mu \rho \alpha} +\bar{p}_\mu\epsilon_{\nu \rho \alpha \beta} \big) 
\bar{p}^\rho n^\alpha  \delta (\bar{p}^2) \bar{D}^\beta  f_\chi^{(0)}  \nonumber \\
&=& \frac{-1}{\bar{p}\cdot n} \left(
 \bar{p}^2 \epsilon_{ \alpha \beta \nu \mu} n^\alpha
+ \bar{p}\cdot n\epsilon_{ \beta \nu \mu \rho } \bar{p}^\rho \right)
  \delta (\bar{p}^2) \bar{D}^\beta  f_\chi^{(0)}
   \nonumber \\
    &=&  \epsilon_{ \mu \nu \beta  \rho } \bar{p}^\rho   \delta (\bar{p}^2) \bar{D}^\beta  f_\chi^{(0)}  \nonumber
\end{eqnarray}
Therefore, we conclude that the semiclassical solution of   (\ref{qke1}) and (\ref{qke3}) is
\begin{equation}
\label{j1}
	\mathcal{J}_{\chi}^\mu= \bar{p}^{\mu} f_\chi \delta (\bar{p}^2) +\hbar q \chi \tilde{\bar{F}}^{\mu\nu} \bar{p}_\nu f_\chi^{(0)} \delta^\prime (\bar{p}^2) + \frac{\hbar \chi}{2\bar{p}\cdot n}\epsilon^{\mu \nu \alpha \beta}
	\bar{p}_\nu n_\alpha\delta (\bar{p}^2)  \bar{D}_\beta  f_\chi^{(0)} 	,
\end{equation}
where we defined the distribution function $ f_\chi = f_\chi^{(0)} +\hbar f_\chi^{(1)}. $ Obviously, it should also satisfy  (\ref{qke2}). We will come back to it in Sec. \ref{kinEq}. Now, we will specify  the equilibrium distribution function and calculate chiral currents.

\section{Fermi-Dirac  distribution and the chiral  currents}
\label{FDD}

The equilibrium distribution function of the particles  which obey 
Fermi-Dirac statistics in the LSV background is studied in \cite{Potting}.
There, the standard relativistic kinetic theory formulation of gases \cite{RBE-CK}  is employed by incorporating the LSV modifications of the equations of motion:
The massless particles obey  the extended dispersion relation 
\begin{equation}
\label{disp}
\bar{p}\cdot \bar{p}=0,
\end{equation}
and the constraint  
\begin{equation}	
\label{const}
	 G^{-1}_{\rho \mu }  u^\mu \left(G^{-1}\right)^{\rho \nu} u_\nu -1=0.
 \end{equation}
 $u_\mu \equiv d x_\mu / d \tau$ is the four-velocity of the fluid where $\tau $ is the proper time in the absence of LSV.
By introducing 
\begin{equation}
	\tilde{u}_\mu\equiv G^{-1}_{\mu\nu} u^\nu, 
\end{equation}
(\ref{const}) can be expressed as
\begin{equation}
	\tilde{u}_\mu\tilde{u}^\mu =1.
\end{equation}

The  Boltzmann and relativistic Uehling-Uhlenbeck equations have been  derived following the ordinary relativistic theory
in terms of  the one-particle phase space distribution function $f(x,p) ,$ and    $\tilde{p}^\mu =G^{\mu\nu} \bar{p}_\nu. $
Then,  the H-theorem is demonstrated by making use of the extended transport equations  and introducing the entropy-density  four-current
\begin{equation}
	s^\mu =\frac{1}{(2\pi \hbar)^3 }\int \frac{d^3p}{\tilde{p}^0} \tilde{p}^\mu  \left[ \left(f(x,p) -1 \right)\ln \left(1-f(x,p)\right) -f(x,p) \ln f(x,p) \right],
\end{equation}
which yields  the total entropy 
\begin{equation}
\label{totent}
	S=\int d^3x s^0=\frac{1}{(2\pi \hbar)^3 }\int d^3xd^3p\,  \left[ \left(f(x,p) -1 \right)\ln \left(1-f(x,p)\right) -f(x,p) \ln f(x,p) \right].
\end{equation}
 The Boltzmann constant is set $k=1.$ Observe that (\ref{totent}) is independent of the LSV coefficients. Thus, as in the ordinary case
 one can observe  that 
for particles with momentum $p_1,p_2$ scattered to particles with momentum $p_3,p_4,$   the total entropy (\ref{totent}) is stationary when $\phi (x,p)=-\ln \left[f(x,p) / (1-f(x,p))\right]$ satisfies the condition
\begin{equation}
	\label{sta}
	\phi (x,p_3 )+\phi (x,p_4 )- \phi ( x,p_1 )-\phi (x,p_2 )=0.
\end{equation}
For momentum  conserving   scatterings, the most general solution of (\ref{sta})  can be shown to be
\begin{equation}
	\label{eqf}
	f (x,p) = \frac{1}{e^{-\alpha (x) +\beta (x) \cdot p } +1},
\end{equation}
where $\alpha (x)$ and $\beta_\mu (x)$ are arbitrary. Equation (\ref{eqf})  should obey the  Uehling-Uhlenbeck equation which is solved by the Fermi-Dirac distribution function 
\begin{equation}
	f_{FD}(x,p) = \frac{1}{e^{(-\mu +u \cdot p )/T} +1},
\end{equation}
for $\partial u_\mu / \partial x^\nu =0.$
On mass-shell $\bar{p}^2=0,$ thus 
the equilibrium distribution function 
for chiral fermions and antifermions  is given by
\begin{eqnarray}
	f_{\chi}^{ eq} & =&\frac{2}{(2\pi \hbar)^3 }\left[\frac{\theta (\tilde{u}\cdot \bar{p})}{e^{(u \cdot p -\mu_\chi)/T} +1}
	+ \frac{\theta (-\tilde{u}\cdot \bar{p})}{e^{-(u \cdot p -\mu_\chi )/T} +1} \right] . \label{dFD1}
\end{eqnarray}
Now by expressing it in $\bar{p}$ we get
\begin{eqnarray}
	f_{\chi}^{ eq} & =&\frac{2}{(2\pi \hbar)^3 }\left[\frac{\theta (\tilde{u}\cdot \bar{p})}{e^{(\tilde{u} \cdot \bar{p} -\mu_\chi -\tilde{u}\cdot a)/T} +1}
	+ \frac{\theta (-\tilde{u}\cdot \bar{p})}{e^{-(\tilde{u}\cdot \bar{p} -\mu_\chi -\tilde{u} \cdot a)/T} +1} \right] . \label{dFD}
\end{eqnarray}
$\mu_\chi $ are given by the total chemical potential $\mu$ and the chiral chemical potential $\mu_5$ as $\mu_{\ssR ,\ssL}=\mu \pm \mu_5.$ Observe that the chemical potentials  $\mu_\chi $ are  effectively shifted with $\tilde{u} \cdot a,$ reminiscent of expressing the distribution function (\ref{dFD1}) in terms of $\bar{p}_\mu,$ which is the LSV extended momentum appearing in the dispersion relation (\ref{disp}).

We would like to study the chiral vector field  (\ref{j1}) by choosing $f_\chi$ as the modified Fermi-Dirac distribution function (\ref{dFD}) and setting $n=\tilde{u}.$ In this frame  we  introduce  $	\bar{E}_\mu =\bar{F}_{\mu\nu} \tilde{u}^\nu ,$ and  $\bar{B}_\mu=(1/2) \epsilon_{\mu \nu \alpha \beta} \tilde{u}^\nu \bar{F}^{\alpha \beta} ,$  so that
the field strength and its dual can be expressed as
%for vanishing vorticity $\omega_{\mu\nu}=0,$ or
%\begin{equation}
	%\frac{\partial u_\mu}{\partial x_\nu}=0,\  \   \  \  	\frac{\partial \theta_\mu}{\partial x_\nu}=0,
%\end{equation}
%\begin{eqnarray}f_{\chi}^{ eq} & =& \frac{2}{(2\pi \hbar)^3 }\left[\frac{\theta ( \bar{p}_0)}{e^{-\alpha+\theta \cdot p } +1}+ \frac{\theta (- \bar{p}_0)}{e^{-\theta\cdot p +\alpha } +1} \right]\\ & =&\frac{2}{(2\pi \hbar)^3 }\left[\frac{\theta (\bar{p}_0)}{e^{-\alpha-\tilde{\theta} \cdot a +\tilde{\theta} \cdot \bar{p} } +1}+ \frac{\theta (-\bar{p}_0)}{e^{-\tilde{\theta}\cdot \bar{p} +\alpha +\tilde{\theta} \cdot a} +1} \right] .\end{eqnarray}Here\begin{equation}\alpha=\frac{\mu_\chi}{kT},\ \ \ \theta_\mu= u_\mu /kT,\ \ \ \ \tilde{\theta}_\mu ={G^{-1}}_\mu^\nu \theta_\nu =\tilde{u}_\mu /kT\end{equation}
\begin{eqnarray}
%	\bar{E}_\mu =\bar{F}_{\mu\nu} \tilde{u}^\nu , \  \  \  \
%		\bar{B}_\mu =\tilde{\bar{F}}_{\mu\nu} \tilde{u}^\nu ; \\
\bar{F}^{\mu\nu} = 	\bar{E}^\mu \tilde{u}^\nu - \bar{E}^\nu \tilde{u}^\mu + \epsilon^{\mu \nu \alpha \beta} \tilde{u}_\alpha \bar{B}_\beta ,\\
\tilde{\bar{F}}^{\mu\nu} = 	\bar{B}^\mu \tilde{u}^\nu - \bar{B}^\nu \tilde{u}^\mu + \epsilon^{\mu \nu \alpha \beta} \tilde{u}_\beta \bar{E}_\alpha  .
\end{eqnarray}

At the zeroth order in Planck constant $f^{eq}_\chi $ should satisfy the LSV Vlasov equation (\ref{Vlasov}), which  for constant temperature $T,$ leads to
\begin{eqnarray}
	G^{\nu\mu}\left(\partial_\mu -qF_{\mu\alpha} \partial_{p}^\alpha \right) f_{\chi}^{ eq} %&=& 	G^\mu_\nu\left(\partial^\nu\mu_\chi+F^{\nu\alpha} \partial_{p\alpha}\bar{p}_0\right)\frac{\partial  f_{\chi}^{ eq}}{\partial \mu_\chi} \nonumber \\	&=& 	G^\mu_\nu\left(\partial^\nu\mu_\chi+F^{\nu\alpha}G_\alpha^0  \right)\frac{\partial  f_{\chi}^{ eq}}{\partial \mu_\chi} \nonumber \\	&=& 	G^\mu_\nu\left(\partial^\nu\mu_\chi+F^{\nu\alpha}G_\alpha^\lambda \tilde{u}_\lambda  \right)\frac{\partial  f_{\chi}^{ eq}}{\partial \mu_\chi} \nonumber \\
	&=& \left(\bar{\partial}^\nu\mu_\chi+q\bar{F}^{\nu\alpha} \tilde{u}_\alpha \right)\frac{\partial  f_{\chi}^{ eq}}{\partial \mu_\chi} =0. \nonumber 		
\end{eqnarray}
It is satisfied by letting
 $\mu=(\mu_\ssR+\mu_\ssL )/2$ and $\mu_5=(\mu_\ssR-\mu_\ssL )/2,$ to fulfil the conditions
\begin{eqnarray}
	\bar{\partial}_{\sigma} \mu=-\bar{E}_\sigma , \ \ \ \   & 	\bar{\partial}_{\sigma} \mu_5= 0. & \label{delmu}
\end{eqnarray}
Observe that $\bar{E}_\sigma =G_{\sigma \mu}E^\mu$ where $E_\mu$ is the electric field, hence, (\ref{delmu}) coincides with $\partial_{\sigma} \mu=-E_\sigma ,\ \partial_{\sigma} \mu_5= 0,$ which are the  relations obeyed in the ordinary chiral kinetic theory  \cite{glpww}. 

The chiral vector field can be expressed as
\begin{eqnarray}
	\mathcal{J}^{eq}_{\chi\mu} &= & \bar{p}_{\mu} f^{eq}_\chi \delta (\bar{p}^2) +\hbar q \chi \tilde{\bar{F}}_{\mu\nu} \bar{p}^\nu f_\chi^{eq} \delta^\prime (\bar{p}^2) + \frac{\hbar \chi}{2\bar{p}\cdot \tilde{u}}\epsilon_{\mu \nu \alpha \beta}
	\bar{p}^\nu \tilde{u}^\alpha\delta (\bar{p}^2)  \bar{D}^\beta  f_\chi^{eq}  \nonumber	\\
&= & \bar{p}_{\mu} f^{eq}_\chi \delta (\bar{p}^2) +\hbar q\chi (\bar{B}_\mu \tilde{u}_\nu - \bar{B}_\nu \tilde{u}_\mu ) \bar{p}^\nu f_\chi^{eq} \delta^\prime (\bar{p}^2) +
\hbar q \chi\epsilon_{\mu \nu \alpha \beta} \tilde{u}^\beta \bar{E}^\alpha \bar{p}^\nu f_\chi^{eq} \delta^\prime (\bar{p}^2) \nonumber\\
&&
 +\frac{\hbar \chi}{2\bar{p}\cdot \tilde{u}}\epsilon_{\mu \nu \alpha \beta}
\bar{p}^\nu \tilde{u}^\alpha\delta (\bar{p}^2) ( \bar{\partial}^\beta \tilde{u}^\kappa) (\bar{p}_\kappa +a_\kappa )\frac{\partial f_\chi^{eq}}{\partial (\tilde{u}\cdot\bar{p}-\tilde{u}\cdot a)}  , \label{JFD}
\end{eqnarray}
where the last term can  also be written as follows
$$
\frac{\hbar \chi}{2\bar{p}\cdot \tilde{u}}\delta (\bar{p}^2)  \epsilon_{\mu \nu \alpha \beta}
\bar{p}^\nu \tilde{u}^\alpha( \bar{\partial}^\beta u^\kappa )p_\kappa \frac{\partial f_\chi^{eq}}{\partial (u\cdot p )}  \cdot
$$

Let us deal with vanishing vorticity\ $\partial^\beta u^\kappa =0 $  and constant electromagnetic fields $E_\mu,B_\mu .$ Then, by employing the identity   $\bar{p}^2\delta^{\prime \prime} (\bar{p}^2 ) =-2\delta^{\prime} (\bar{p}^2)$ and the relation
\begin{equation}
\epsilon^{\mu \nu \rho \beta} \epsilon_{\beta \sigma  \delta \alpha } =
   \delta^\mu_\sigma\left(\delta_\delta^\nu \delta_\alpha^\rho - \delta_\delta^\rho \delta_\alpha^\nu\right) 
+ \delta^\nu_\sigma\left(\delta_\delta^\rho \delta_\alpha^\mu - \delta_\delta^\mu \delta_\alpha^\rho\right) 
+\delta^\rho_\sigma\left(\delta_\delta^\mu \delta_\alpha^\nu - \delta_\delta^\nu \delta_\alpha^\mu\right) ,
\end{equation}  
one can easily 
observe that (\ref{JFD}) satisfies the remaining equation (\ref{qke2}).

Chiral current is defined as
\begin{equation}
	j^\mu_\chi (x)=\int d^4p\, 	\mathcal{J}^{\mu}_{\chi} (x,p).
\end{equation}
By the change of  variables $p_\mu \rightarrow \bar{p}_\mu ,$ we get
\begin{equation}
	j^\mu_\chi (x)=\frac{1}{|G|}\int d^4\bar{p}\, 	\mathcal{J}^{\mu}_{\chi} (x,p),
\end{equation}
where $|G|=\det G^\mu_\nu .$

%Define $\bar{p}^\mu=(\bar{p}\cdot \tilde{u} )\tilde{u}^\mu +\bar{p}_\perp^\mu .$ 
To perform the integrals over $\bar{p}^\mu =(\bar{p}^0, \vec{\bar{p}}),$  let us designate $|\vec{\bar{p}}|\equiv \bar{P}$ and note  that
\begin{equation}
	\int_0^\infty d\bar{P}\, \bar{P}^k \frac{1}{e^{\left[\bar{P} -\mu_\chi -\tilde{u}\cdot a\right]/T} +1}
=-T^{k+1} k!\,  Li_{k+1} \left(-e^{(\mu_\chi +\tilde{u}\cdot a)/T}\right),
\end{equation}
where  $Li_s(x)$ are the polylogarithms  whose properties which we use in the calculations can be found in \cite{dk2017}.

By performing the integrals we get
\begin{equation}
		j_{\chi}^\mu =n_\chi \tilde{u}^\mu +\xi_\chi  \bar{B}^\mu,
\end{equation}
with
\begin{eqnarray}
	n_\chi & = & \frac{1}{6\pi^2\hbar^3|G|} \left[\left(\mu_\chi +\tilde{u}\cdot a \right)^3+\pi^2 T^2 \left(\mu_\chi + \tilde{u}\cdot a\right)\right] , \\
	\xi_\chi & = &   \frac{1}{4\pi^2\hbar^2|G|}  \left(\mu_\chi + \tilde{u}\cdot a\right) .
\end{eqnarray} 
Then, the vector and axial-vector currents defined by $j^\mu=j^\mu_R+j^\mu_L$ and  $j^\mu_5=j^\mu_R-j^\mu_L$ are established as
\begin{eqnarray}
	j^\mu &=& n \tilde{u}^\mu +q\xi  \bar{B}^\mu, \label{cMe}\\
		j^\mu_5 &=& n_5 \tilde{u}^\mu +q\xi_5  \bar{B}^\mu , \label{cSe}
\end{eqnarray}
where
\begin{eqnarray}
	n & = & \frac{1}{3\pi^2\hbar^3|G|} \left[\mu^3 +(\tilde{u}\cdot a )^3\right] + \frac{1}{\pi^2\hbar^3|G|}\left[\mu \mu^2_5 +(\mu^2+\mu_5^2)(\tilde{u}\cdot a  )+\mu (\tilde{u}\cdot a  )^2\right] \nonumber \\
	&& +\frac{T^2}{3\hbar^3|G|} (\mu+\tilde{u}\cdot a  ) \\
		n_5 & = & \frac{1}{6\pi^2\hbar^3|G|} \left[\left(\mu_\chi +\tilde{u}\cdot a \right)^3+\pi^2 T^2 \left(\mu_\chi + \tilde{u}\cdot a\right)\right] , \\
			\xi& = &   \frac{\mu_5}{2\pi^2\hbar^2|G|} ,\label{xi}\\
	\xi_5& = &   \frac{\mu + \tilde{u}\cdot a}{2\pi^2\hbar^2|G|}  \label{xi5}
\end{eqnarray} 
In the absence of LSV terms, the magnetic parts of the currents (\ref{cMe}) and (\ref{cSe})  are known as chiral magnetic effect (CME) and chiral separation effect (CSE). 
 We found that these effects are modified in two ways. First, the magnetic field is replaced by $\bar{B}$, whose dependence on the magnetic field 
 $B_\mu=(1/2) \epsilon_{\mu \nu \alpha \beta} u^\nu F^{\alpha \beta} $ is not direct. Second,  the coefficients  (\ref{xi}) and (\ref{xi5}) depend on the LSV terms.

By making use of (\ref{delmu}) one observes that the vector current is conversed but the axial-vector current is anomalous:
\begin{eqnarray}
	\partial_{\mu} j^\mu &=&0 , \\
	 \partial_{\mu} j_5^\mu &=& \frac{q^2}{2\pi^2\hbar^2|G|} E\cdot \bar{B} \label{anom} .
\end{eqnarray}
The chiral anomaly in this context has been studied in \cite{Sch,Scar}. In \cite{Sch} $a=0,$ and  the chiral current was defined as $G_{\mu \nu }J^\nu_5$ where $J_5$ is the ordinary axial-vector current. Thus, its magnetic part is similar to (\ref{cMe}). By studying the index theorem, they concluded that the chiral anomaly is the same as in the ordinary Lorentz-invariant theory. The same conclusion is drawn in  \cite{salv}.   However, in \cite{Scar}, the  chiral anomaly was calculated  by employing the Fujikawa method as 
\begin{eqnarray}
\bar{\partial}_{\mu} j_5^\mu &=& \frac{q^2}{2\pi^2\hbar^2|G|} \bar{E}\cdot \bar{B} . \nonumber
\end{eqnarray}
This coincides with our result  (\ref{anom}).

\section{Chiral transport equation}
\label{kinEq}

As we have already mentioned, (\ref{j1}) should  also  satisfy (\ref{qke2}):
\begin{equation}
	\bar{D}\cdot	\mathcal{J}_{\chi} 
= \bar{D}^{\mu} \Big[\bar{p}_{\mu} f_\chi \delta (\bar{p}^2) +\hbar q\chi \tilde{\bar{F}}_{\mu\beta} \bar{p}^\beta f_\chi^{(0)} \delta^\prime (\bar{p}^2) + \frac{\hbar \chi}{2\bar{p}\cdot n}\epsilon_{\mu \nu \alpha \beta}
\bar{p}^\nu n^\alpha\delta (\bar{p}^2)  \bar{D}^\beta  f_\chi^{(0)} 	\Big]
	=0.\label{D11}
\end{equation}
Following the approach in \cite{hsjlz}, we will show that (\ref{D11})  leads to the LSV extended chiral transport equation. 
 Since $\bar{F}_{\mu\nu}$ is antisymmetric, the first term can easily be shown to be
\begin{equation}
\bar{D}^{\mu} \big[\bar{p}_{\mu} f_\chi \delta (\bar{p}^2) \big]	=  \delta (\bar{p}^2) \bar{p}_{\mu} \bar{D}^{\mu} f_\chi  .\label{1t}
\end{equation}
Let us focus on  the second term of (\ref{D11}). First, note that  
\begin{equation}
4  \tilde{\bar{F}}^{\mu\beta} \bar{p}_\beta  \bar{F}_{\mu\nu} \bar{p}^\nu = \tilde{\bar{F}}^{\mu\beta} \bar{F}_{\mu\beta} \bar{p}^2 ,
\end{equation}
which is derived by means of the Schouten identity (\ref{sid}).
 By employing the identity  (\ref{sid}) and 
 the relations $\bar{p}^2\delta^{ \prime} (\bar{p}^2 ) =-\delta (\bar{p}^2),$  $\bar{p}^2\delta^{\prime \prime} (\bar{p}^2 ) =-2\delta^{\prime} (\bar{p}^2),$  
 we get
\begin{equation}
\bar{D}^{\mu} \big[ \tilde{\bar{F}}_{\mu\beta} \bar{p}^\beta f_\chi^{(0)} \delta^\prime (\bar{p}^2) \big]=  \tilde{\bar{F}}_{\mu\beta} \bar{p}^\beta  \delta^\prime (\bar{p}^2) 
\bar{D}^{\mu} f_\chi^{(0)}. \label{t2}
\end{equation}
The third term of (\ref{D11})  can be expressed as
\begin{eqnarray}
 \bar{D}^{\mu} \Big[\frac{1}{2\bar{p}\cdot n}\epsilon_{\mu \nu \alpha \beta}
	\bar{p}^\nu n^\alpha\delta (\bar{p}^2)  \bar{D}^\beta  f_\chi^{(0)} 	\Big] = - q\tilde{\bar{F}}^{\mu\nu}\bar{p}_\nu   \delta^\prime (\bar{p}^2) \bar{D}_\mu  f_\chi^{(0)} -\frac{q}{\bar{p}\cdot n} \bar{p}_\mu \tilde{\bar{F}}^{\mu\nu} n_\nu  \delta^\prime (\bar{p}^2)\bar{p}\cdot \bar{D} f_\chi^{(0)}
	 \nonumber \\
	 -\frac{q}{2(\bar{p}\cdot n)^2}\epsilon_{\mu \nu \alpha \beta} \bar{F}^{\mu\rho} n_\rho n^\nu \bar{p}^\alpha \delta (\bar{p}^2) \bar{D}^\beta f_\chi^{(0)} \nonumber \\
+\frac{q}{4\bar{p}\cdot n}\epsilon_{\mu \nu \alpha \beta} \bar{p}^\nu n^\alpha\delta (\bar{p}^2)\big[ G^\beta_\rho (\bar{\partial}^\mu F^{\rho \sigma}) - G^\mu_\rho (\bar{\partial}^\beta F^{\rho \sigma})\big]  \partial_{p \sigma}  f_\chi^{(0)}  \nonumber \\
 - \frac{1}{2(\bar{p}\cdot n)^2}\epsilon_{\mu \nu \alpha \beta}  (\bar{\partial}^\mu n^\rho)\bar{p}_\rho	
 n^\nu \bar{p}^\alpha \delta (\bar{p}^2) \bar{D}^\beta f_\chi^{(0)}  + \frac{1}{2\bar{p}\cdot n}\epsilon_{\mu \nu \alpha \beta}(\bar{\partial}^\mu n^\nu)\bar{p}^\alpha
  \delta (\bar{p}^2) \bar{D}^\beta f_\chi^{(0)} .\  \label{t3}
\end{eqnarray}
Here, we utilized the following equality, derived by making use of the Schouten identity (\ref{sid}), 
\begin{eqnarray}
	\epsilon^{ \mu \nu \lambda  \rho } \bar{F}_{\mu\sigma} \bar{p}^\sigma n_\nu \bar{p}_\lambda \delta^\prime (\bar{p}^2) \bar{D}_\rho  f_\chi^{(0)} 
= (\bar{p}\cdot n)\tilde{\bar{F}}^{\lambda\rho} \bar{p}_\lambda \delta^\prime (\bar{p}^2) \bar{D}_\rho  f_\chi^{(0)} \nonumber \\
+n_\nu  \tilde{\bar{F}}^{\nu\lambda} \bar{p}_\lambda  \bar{p}^\rho \delta^\prime (\bar{p}^2) \bar{D}_\rho  f_\chi^{(0)} 
+ \tilde{\bar{F}}^{\rho \nu} n_\nu\bar{p}^2 \delta^\prime (\bar{p}^2) \bar{D}_\rho  f_\chi^{(0)} ,
\end{eqnarray}
and the commutator relation
\begin{eqnarray}
\epsilon^{\mu \nu \lambda \rho } \left[\bar{D}_\mu , \bar{D}_\rho \right]  f_\chi^{(0)}  =q	\epsilon^{\mu \nu \lambda \rho } G_\mu^\alpha  G_\rho^\beta \left[\left(\partial_\alpha F_{\beta \sigma} \right) - \left(\partial_\beta F_{\alpha \sigma} \right)
 \right]  \partial_{p}^\sigma  f_\chi^{(0)} .
\end{eqnarray}
Now, by inserting (\ref{1t}), (\ref{t2}) and (\ref{t3}) into (\ref{D11}), one establishes
\begin{eqnarray}
	\delta (\bar{p}^2 	-\frac{\hbar q\chi}{\bar{p}\cdot n} \bar{p}_\mu \tilde{\bar{F}}^{\mu\nu} n_\nu  ) \Big\{\bar{p} \cdot \bar{D} f_\chi  
	-\frac{\hbar q \chi}{2(\bar{p}\cdot n)^2}\epsilon_{\mu \nu \alpha \beta} \bar{F}^{\mu\rho} n_\rho n^\nu \bar{p}^\alpha \delta (\bar{p}^2) \bar{D}^\beta f_\chi^{(0)} \nonumber \\
	+\frac{\hbar q\chi}{4\bar{p}\cdot n}\epsilon_{\mu \nu \alpha \beta} \bar{p}^\nu n^\alpha\delta (\bar{p}^2)\big[ G^\beta_\rho (\bar{\partial}^\mu F^{\rho \sigma}) - G^\mu_\rho (\bar{\partial}^\beta F^{\rho \sigma})\big]  \partial_{p \sigma}  f_\chi^{(0)}  \nonumber \\
	- \frac{\hbar \chi}{2(\bar{p}\cdot n)^2}\epsilon_{\mu \nu \alpha \beta}  (\bar{\partial}^\mu n^\rho)\bar{p}_\rho	
	n^\nu \bar{p}^\alpha \delta (\bar{p}^2) \bar{D}^\beta f_\chi^{(0)}  + \frac{\hbar \chi}{2\bar{p}\cdot n}\epsilon_{\mu \nu \alpha \beta}(\bar{\partial}^\mu n^\nu)\bar{p}^\alpha
	\delta (\bar{p}^2) \bar{D}^\beta f_\chi^{(0)} \Big\}=0. \label{kte}
\end{eqnarray}
Obviously, one should keep only the terms up to the $\hbar$ order. Equation (\ref{kte})  is the semiclassical chiral  transport equation, which is observer Lorentz invariant.
The $\delta$ function dictates the mass-shell condition. 

\section{Discussions}

 We studied the fermionic part of the extended QED which is  the modified Dirac equation with a set of LSV terms, in the presence of external electromagnetic fields. It is  observer Lorentz symmetric as is reminiscent of the spontaneously broken Lorentz invariance of  the  fundamental theory which may be string theory.
We proceed by quantizing spinor fields using standard methods and utilizing these spinor operators to construct a relativistic Wigner function through conventional quantum field theory techniques. 

The quantum kinetic equation governing  the Wigner function is accomplished by  following the well-established  formulation of \cite{vge}. 
By decomposing  the Wigner function in terms of the Clifford algebra generators constructed from the $\gamma$ matrices,  we derive a set of coupled equations satisfied by  the scalar, pseudoscalar, vector, axial-vector and antisymmetric tensor fields. Notably, in the   massless limit, the equations of the vector and axial-vector fields are decoupled from the rest. In fact, we focus on  the chiral fermions and adopt the semiclassical approximation where the terms up to the first order in Planck constant are considered. 

Within this  semiclassical approximation we solve  two of the three kinetic equations of the chiral vector fields.  Subsequently, we compute particle four-currents by adopting the analog of the Dirac-Fermi distribution  proposed in \cite{Potting}
The extended magnetic and separation effects are established. The vector current is conserved and the axial-vector current is anomalous, where the chiral anomaly depends on the LSV coefficients. Then, by imposing the third equation which should be satisfied by the chiral vector fields,  the chiral semiclassical kinetic equation is accomplished.
 
How can one experimentally test the results of the modified chiral kinetic theory? This may be achieved through the currents (\ref{cMe}) and (\ref{cSe}), which yield the CME and CSE. In high-energy physics, experimental evidence for the CME is investigated in heavy-ion collisions. The current experimental status has been recently reviewed in \cite{KLT}.
We have seen that LSV parameters modify the coefficient of the magnetic field (\ref{xi}), by $|G|^{-1},$ which would be nearly impossible to detect.  However, note that the modified current is along $\bar{\bm B},$ hence the direction of the CME is altered by LSV parameters. Although this effect is small, it is the unique source in the deviation in the direction. This can even help in experimentally observing  the CME, which has been investigated only along the direction of the magnetic field.
 
In principle, one could derive semiclassical kinetic equations for massive fermions by examining the defining equations (\ref{eq1})-(\ref{eq4}). However, even in the case of ordinary Dirac fermions, this task is intricate, as evidenced by the complexities encountered when employing the approaches outlined in \cite{wsswr,hhy}.

Integration of (\ref{kte}) over the zeroth component of momentum to obtain the nonrelativistic (3D) chiral kinetic theory is a desired step, albeit complicated by the presence of $G_{\mu \nu}$. A pragmatic solution involves setting $a_0=0,\ G_{0\mu} =\delta_{0\mu}$, yielding $\bar{p}_0 =p_0,$  which facilitates to get the mass-shell condition  dictated by the Dirac $\delta$ function. 

By integrating the equations governing the components of the Wigner function (\ref{eq1})-(\ref{eq4}), with respect to the zeroth component of momentum, one can derive what is known as the equal-time formulation, a technique elucidated in detail in \cite{oh, zhuang96-rke, Wang_2021,Efe}.

While our study primarily focuses on establishing chiral currents under conditions of vanishing vorticity, the presence of vorticity may necessitate modifications to the theory, akin to conventional scenarios as discussed in \cite{dk, dk21}.

%\cite{dk2017}

%\cite{dk}
%	\newpage
%\bibliographystyle{apsrev4-2}
\bibliography{SMEWR}

%\bibliography{SMEWR.bib}
\end{document}